\documentclass[showpacs]{revtex4}
\usepackage{latexsym}

\begin{document}

\title{Hydrostatic equilibrium equation and Newtonian limit of the singular $f(R)$ gravity}

\author{A. J. Bustelo and D.E.
Barraco}
 \affiliation{FaM.A.F., Universidad Nacional
de C\'ordoba \\
Ciudad Universitaria, C\'ordoba 5000, Argentina} \email{barraco \@
famaf.unc.edu.ar}

\begin{abstract}
We derive the hydrostatic equilibrium equation of a spherical star
for any gravitational Lagrangian density of the form
$L=\sqrt{-g}f(R)$. The Palatini variational principle for the
Helmholtz Lagrangian in the Einstein gauge is used to obtain the
field equations in this gauge. The equilibrium hydrostatic equation
is obtained and is used to study the Newtonian limit for
$f(R)=R-\frac{a^{2}}{3R}$. The same procedure is carried out for the
more generally case $f(R)=R-\frac{1}{n+2}\frac{a^{n+1}}{R^{n}}$
giving a good Newtonian limit.
\end{abstract}

\maketitle
%------------------------------------------------------------------------------------------------------------
\section*{I. Introduction}
From recent studies it seems well established that our universe is
currently in an accelerating phase. The evidence of cosmic
acceleration has arisen not only from the high-redshift surveys of
type Ia supernovae \cite{Ri98.1,Pe99.1}, but also from the
anisotropy power spectrum of the cosmic microwave background
\cite{Be03.1,Ne02.1}. One of the most accepted explanations is that
the universe has been dominated by some form of dark energy for a
long time. However, none of the existing dark energy models are
completely satisfactory.

It is possible to find other explanations for cosmic expansion using
field equations other than Einstein's equations. Recently, some
authors have proposed to add a $R^{-1}$ term in the Einstein-Hilbert
action to modify general relativity \cite{Ca02.1}. They have
obtained the field equations using the second order formalism,
varying only the metric field, and have thus derived the so-called
fourth-order field equations. Although the models were obtained
using corrections of the Hilbert-Einstein Lagrangian of type
$R^{n}$, where $n$ can take a positive or negative value to explain
both the inflation at an early time and the expansion at the present
time\cite{No03.1}. Apart from this \emph{ad hoc} justification,
there are also theoretical motivations for non-linear gravity from
M-theory \cite{No03.1}. However, this model still suffer violent
instabilities \cite{Do03.1} which maybe deleted by the addition of a
$R^2$-like term to the Lagranngian \cite{No03.1} or by account of
quantum effects \cite{Do6.10.06}. A big review on various modified
gravities and its applications to cosmology was analyzed in
\cite{Od06.1}, and the Newtonian limit of these fourth order
theories has been studied by Dick \cite{Di04.1}

There are also theories that are obtained from a Lagrangian density
$L_{t}=\sqrt{-g}f(r)+L_{m}$, which depends on the scalar of
curvature and a matter Lagrangian does not depends on the
connection,on this Lagrangian we can apply Palatini's method to
obtain the field equations \cite{Ha93.1,Barr91.1}. In references
\cite{Ha93.1,Barr91.1}, we showed the universality of the Einstein
equation using a cosmological constant. More recently, M. Ferraris,
M.Francaviglia and I. Volovich  published the same result
\cite{Fran93.1, Fran94.1}. For these theories we have studied the
conserved quantities \cite{Barr99.1}, the spherically symmetric
solutions \cite{Barr00.1}, the Newtonian limit
\cite{Ha93.1,Barr91.1,Barr96.1}, and the Cosmology described by FRW
metric \cite{Barr02.1}.

In \cite{Barr96.1}, it was shown that is very difficult to test these
models in the (post-) Newtonian approximation. The reason for this
is that the departures from Newtonian behavior are both very small
and are masked by other effects, due to the fact that these
departures have to be measured when the body is moving "through" a
matter-filled region.

Recently, Vollick \cite{Vo03.1} used the Lagrangian
$f(R)=R-\frac{a^{2}}{3R}$, together with the Palatini variational
principle to prove that the solution of field equations approach the
De Sitter universe at a late time. This result was obtained using,
the well known property of the vacuum solutions of these theories,
that in a vacuum, or in the case of $T=constant$, the solutions are
the same as in General Relativity with a cosmological constant, even
when $ f(R)$ is not analytical at
$R=0$\cite{Barr96.1,Fran94.1,Barr00.1}.

On the other hand, solutions corresponding to different cosmological
constants are allowed by some of these theories. Therefore, the
homogeneous and isotropic vacuum solution for these theories is the
deSitter space-time with different cosmological constants, except
when one of the  allowed cosmological constants is $\Lambda=0$,
which corresponds to flat space-time.
 Thus, the inclusion
of $1/R$ curvature terms in the gravitational Lagrangian provides us
with an alternative explanation for cosmological acceleration. The
generalization to the case of scalar tensorial theories was analyzed
in \cite{Od06}.

 In the present work, we use the first order or Palatini
formalism \cite{Ha93.1} in order to obtain the hydrostatic
equilibrium equation, for theories that are obtained from the
Lagrangian density $L=\sqrt{-g}f(R)$ which depends on the scalar of
curvature, and a matter Lagrangian that does not depend on the
connection. This is done by means of a Legendre transformation which
in classical mechanics replaces the Lagrangian of a mechanical
system with the Helmholtz Lagrangian \cite{Ma94.1}. From this
equation we have found the Newtonian limit in the particular case $
f(R)=R-\frac{a^{2}}{3R}$, is a good Newtonian limit in same cases
and is not in others. This limit has been treated in the literature
 \cite{So06.1,Do04.1}with no consensus about what the background
metric is, that has to be perturbed. In this sense, the difference
from previous works is that we obtain this limit from the
hydrostatic equilibrium equation and we assume nothing about the
background metric. In addition, we generalize the above result for
the case $f(R)=R-\frac{1}{n+2}\frac{a^{n+1}}{R^{n}}$. Also, for this
theory we see that the departure from Newtonian behavior is smaller
as $n$ increase.

 In the next section we review the field equations and
 the basic structure of the theory. In section III we derive
 the static spherically symmetric solution. In section IV we
 derive the equilibrium hydrostatic equation, modeling the star as a
 perfect fluid for $f(R)$ theories of gravity. In the following section
 we determine the weak field limit from this equation. In section VI we
 calculate the weak field limit for singular $f(R)$ gravity model, and
 in section VIII we show the conclusions.
 This work is part of a graduation thesis \cite{bust1}

\section*{II. Helmholtz Lagrangian in the first order formalism}

 The action for an $f$($R$)  gravity in the Jordan(original) gauge
with the metric  ${g}_{ ab}$ is given by \cite{Po06.1}
\begin{equation}\label{eq:1}
    S=\frac{1}{2\alpha_{m}}\int\sqrt{-g}f(R)d^{4}x+S_{m}(g_{
    ab},\psi),
\end{equation}
Here $\sqrt{-g}f(R)$, is a Lagrangian density that depends on  the
curvature scalar $R=R_{ ab}(\Gamma^{c}_{de})g^{ab}$, $S_{m}$ is the
action for matter represented symbolically by $\psi$ and independent
of the connection, and $\alpha_{m}=8\pi G$, where we have taken
$c\equiv1$.

We consider that we have a smooth one-parameter $(\lambda)$  family
of field configurations starting from given fields $g^{ ab}$,
$\Gamma^{c}_{de}$ and  $\psi$, with appropriate boundary conditions
and, denote by $\delta g^{ab}$, $\delta\Gamma^{c}_{ab}$,
$\delta\psi$ the corresponding variations, i.e., $\delta g^{
ab}=\frac{dg^{ ab}}{d\lambda}|_{\lambda=0}\delta\lambda$, etc.
 Variation of the action $S_{j}$
with respect to $g_{ ab}$ yields the field equations \cite{Ha93.1}
\begin{equation}\label{eq:2}
    f'(R)R_{ ab}-\frac{1}{2}f(R)g_{ ab}=\alpha_{m}T_{ ab},
\end{equation}
where $f'(R)=(\frac{df}{dR})$ and the dynamical energy-moment tensor
of matter is generated taking variation of the action of matter with
respect to the Jordan metric tensor:
\begin{equation}\label{eq:3}
    \delta S_{m}=-\frac{1}{2}\int\ T_{ ab}\sqrt{-g} d^{4}x\delta g^{ ab}.
\end{equation}
  The trace of field equation (\ref{eq:2}), gives
\begin{equation}\label{eq:4}
    f'(R)R-2f(R)=\alpha_{m}T,
\end{equation}
where (\ref{eq:4}) fixes a unique relation, in general non-linear,
between $R$ and $T$.
  Varying the action with respect to the connection, and
recalling that this is fixed al the boundary of $\mathcal{U}$ , we
obtain:
\begin{equation}\label{eq:5}
    \nabla^{\Gamma}_{c}(\sqrt{-g}g^{ ab}f'(R))=0,
\end{equation}
 this equation define, the connection $\Gamma$ which is not the
metric connection. The metric connection satisfies the equation
$\nabla_c g_{ab} =0$. From the field equation (\ref{eq:5}), it
follows that the connection $\Gamma$ is given by the Christoffel
symbols of the conformal transformed metric:
\begin{equation}\label{eq:6}
\tilde{g}_{ ab}=f'(R)g_{ ab}.
\end{equation}
The metric $\tilde {g}_{ ab}$ defines the Einstein gauge in which
the connection $\Gamma$ is metric compatible. Therefore, instead of
using $\nabla^{\Gamma}_c$ we will use $\tilde{\nabla}_c$.

The action (\ref{eq:1}) is dynamically equivalent to the Helmholtz
action, given by \cite{Ma94.1}
\begin{equation}\label{eq:7}
    S_{H}= \frac{1}{\alpha_{m}} \int d^{4}x
    \sqrt{-g} [f(\phi(p))+p(R(g)-\phi(p))]+ S_{m}(g_{ ab},\psi),
\end{equation}
where $p$ is a new scalar variable, and $\phi(P)$ is an auxiliary
scalar field determined by the following equation
\begin{equation}\label{eq:8}
    {\frac{\partial f}{\partial R}} |_{R=\phi(p)}=p.
\end{equation}

Taking variations of the action with respect to $p$, and assuming
that the Lagrangian of matter does not depend on the connection, we
have:
\begin{equation}\label{eq:9}
    \phi(p)=R(g_{ ab}).
\end{equation}
 From the form of the scalar of curvature in the Jordan gauge and
 from (\ref{eq:6}) we obtain
 \begin{equation}\label{eq:10}
    \phi=\tilde{R}f'(\phi)=\tilde{R}p,
 \end{equation}
where $\tilde{R}=R_{ ab}(\Gamma^{c}_{ de})\tilde{g}^{ ab}$ is the
scalar of curvature of the metric $\tilde{g}_{ ab}$.

Transforming (\ref{eq:7}) to the Einstein gauge this action becomes
the standard Hilbert-Einstein action with an additional scalar field
\begin{equation}\label{eq:11}
    S_{E}=\frac{1}{2\alpha_{m}}\int
    d^{4}x\sqrt{-\tilde{g}}[\tilde{R}-\frac{\phi(p)}{p}+\frac{f(\phi(p))}{p^{2}}]+S_{m}(p^{-1}\tilde{g}_{
    ab},\psi)],
\end{equation}
and choosing $\phi$ as the scalar variable (note that $\phi(p)=R_{
ab}(\Gamma)g^{ ab}$ i.e. is the scalar of curvature in the Jordan
gauge) leads to
\begin{equation}\label{eq:12}
    S_{E}=\frac{1}{2\alpha_{m}}\int
    d^{4}x\sqrt{-g}[\tilde{R}-2V(\phi)] +S_{m}(\tilde{g}_{
    ab}[f'(\phi)]^{-1},\psi)
\end{equation}
where $V(\phi)$ is the potential:
\begin{equation}\label{eq:13}
    V(\phi)=\frac{\phi f'(\phi)-f(\phi)}{2(f'(\phi))^{2}}.
\end{equation}

 Now the variation of the action with respect to $\tilde{g}_{
 ab}$ gives the field equations of the action (\ref{eq:1}) in the
 Einstein  gauge \cite{Po06.1}:
 \begin{equation}\label{eq:14}
    \tilde{R}_{ ab}-\frac{1}{2}\tilde{R} \tilde{g}_{ ab}=\frac{\alpha_{m}T_{ab}([f'(\phi)]^{-1}\tilde{g}_{
    ab})}{f'(\phi)}-V(\phi)\tilde{g}_{ ab}.
 \end{equation}

Then, considering (\ref{eq:10}) and taking trace of the before field
equation we obtain
\begin{equation}\label{eq:15}
    \phi f'(\phi)-2f(\phi)=\alpha_{m}T([f'(\phi)]^{-1}\tilde{g}_{
    ab})f'(\phi),
\end{equation}
with the last equation being equivalent to equation (\ref{eq:4}) but
written in the Einstein gauge.

\section*{III. Static spherically symmetric solution}
For the two cases of $f(R)$ theories, those which emerge from second
order formalism and those which emerge from first order formalism,
the static spherical symmetric solution for vacuum has been studied
in \cite{Od05.1,Barr00.1} Let us work in the Einstein gauge and
consider now a static spherically symmetric spacetime manifold with
metric $\tilde{g}_{ ab}$ in the standard form:
\begin{equation}\label{eq:16}
    d\tilde{s}^2=e^{\alpha(r)}dr^{2}+r^{2}d\omega^{2}-e^{\gamma(r)}dt^{2},
\end{equation}
where the coordinates are curvature coordinates of the metric
$\tilde{g}_{ab}$.

The nonzero components of the Einstein tensor \cite{Sy64.1},
$\tilde{G}_{ ab}=\tilde{R}_{ ab}-\frac{1}{2}\tilde{R}\tilde{g}_{
ab}$, are

    $$\tilde{G}^{r}_{r}=\frac{e^{-\alpha}}{r^{2}}(1+r\gamma_{,r})-\frac{1}{r^{2}},$$

    $$\tilde{G}^{\theta}_{\theta}=\tilde{G}^{\varphi}_{\varphi}=e^{-\alpha}(\frac{\gamma_{, rr}}{2} +
    \frac{\gamma^{2}_{,r}}{4}+\frac{\gamma_{, r}}{2r}-\frac{\alpha_{, r}}{2r}-\frac{\alpha_{, r}\gamma_{,r}}{4}),$$

\begin{equation}\label{eq:17}
    \tilde{G}^{t}_{t}=\frac{e^{-\alpha}(1-r\alpha_{,
    r})}{r^{2}}-\frac{1}{r^{2}}.
\end{equation}

Let us now regard $T_{t}^{t}$ and $T^{r}_{r}$ as assigned then the
solution of the field equations (\ref{eq:14}), in the region $r\leq
r_{0}$ where $r_{0}$ is the  proper radius of the star, are

\begin{equation}\label{eq:18}
    e^{-\alpha}=1-\frac{2\tilde{M}(r)}{r},
\end{equation}

\begin{equation}\label{eq:19}
    e^{\gamma}=e^{-\alpha}+exp(\int^{r}_{0}\hat{r}e^{\alpha}\frac{\alpha_{m}}{f'(\phi)}[T^{r}_{r}-T^{t}_{t}]d\hat{r}),
\end{equation}
where $\tilde{M}(r)$ is given by:

\begin{equation}\label{eq:20}
    \tilde{M}(r)=-\frac{1}{2}\int^{r}_{0}\hat{r}^{2}(\frac{\alpha_{m}T^{t}_{t}}{f'(\phi)}-\frac{¨[\phi
    f'(\phi)-f(\phi)]}{2[f'(\phi)]^{2}})d\hat{r}.
\end{equation}

 The spherical symmetry of the metric is, of course, invariant under
 the conformal change (\ref{eq:6}).Thus, to return
to the original gauge, we have to perform the inverse transformation
(\ref{eq:6}) and we can write the metric $g_{ ab}$ in
 the form

 \begin{equation}\label{eq:21}
ds^{2}=\frac
{e^{\alpha(r)}}{f'(r)}dr^{2}+\frac{r^{2}}{f'(r)}d\omega^{2}-\frac{e^{\gamma
(r)}}{f'(r)}dt^{2}.
 \end{equation}
 It is convenient to put the metric in the standard form in order to know the physical meaning of the
  coordinates, therefore we change the coordinates from the curvature coordinates of the $\tilde{g}_{ab}$
  to the curvature coordinates of the original metric$g_{ab}$.
 In this coordinates $r'=\frac{r}{\sqrt{f'(\phi)}}$ and the original metric $g_{ab}$ is
\begin{equation}\label{eq:21a}
ds^2 = A(r') dr'^2 + r'^2 dw^2 - B(r')dt^2,
\end{equation}
 where \begin{equation}\label{eq:22}
    A(r')=\frac{[\frac{r'}{\sqrt{f'(\phi)}}\frac{d\sqrt{f'(\phi)}}{dr'}+1]^{2}}{1-\frac{2\tilde{M}(r')}{r'}},
\end{equation}
and
\begin{equation}\label{eq:23}
    B(r')=\frac{1}{f'(\phi)}[1-\frac{2\tilde{M}(r')}{r'}+exp
    (\int^{r'\sqrt{f'(\phi)}}_{0}\hat{r}e^{\alpha}\frac{\alpha_{m}}{f'(\phi)}(T^{r}_{r}-T^{t}_{t})d\hat{r})],
\end{equation}
 $\tilde{M}(r')$ is:
\begin{equation}\label{eq:24}
    \tilde{M}(r')=-\frac{1}{2\sqrt{f'(\phi)}}\int^{r'\sqrt{f'(\phi)}}_{0}\hat{r}^{2}(\frac{\alpha_{m}}{f'(\phi)}T^{t}_{t}-\frac{[\phi
    f'(\phi)-f(\phi)]}{2[f'(\phi)]^{2}})d\hat{r}.
\end{equation}
This results are similar to that obtained in \cite{Barr00.1}, in
that work was also proved that in all cases the exterior metrics
match correctly with the interior solutions.

\section*{IV. The hydrostatic equilibrium equation}

 In the original spacetime, Jordan gauge, the energy momentum tensor corresponds to
 a perfect fluid model:
 \begin{equation}\label{eq:25}
    T_{ab}(g_{cd})=\rho u_{a}u_{b}+p(g_{ ab}+u_{a}u_{b}).
 \end{equation}

 In order to obtain $T_{ab}(f'^{-1}\tilde{g}_{cd})$ we note that the
 energy momentum tensor will be of the form of a perfect fluid because $u_{a}u^{a}=\tilde{u}_{a}\tilde{u}^{a}=-1$; i.e. $\tilde{u}_{a}\tilde{u}_{b}=f'(\phi)u_{a}u_{b}$. Thus
   $u_{a}=\frac{1}{\sqrt{f'(\phi)}}\tilde{u}_{a}$ and
\begin{equation}\label{eq:26}
    T_{ab}(\tilde{g}_{cd}[f'(\phi)]^{-1})=\tilde{\rho}\tilde{u}_{a}\tilde{u}_{b}+\tilde{p}(\tilde{g}_{
    ab}+\tilde{u}_{a}\tilde{u}_{b}),
\end{equation}
where $\tilde{p}=p/f'(\phi)$ and $\tilde{\rho}=\rho/f'(\phi)$. In
order to derive the equation of hydrostatic equilibrium the field
equation that we are interesting in, is
\begin{equation}\label{eq:27}
    \tilde{G}^{r}_{r}=\frac{e^{-\alpha}}{r^{2}}(1+r \gamma_{,
    r})-\frac{1}{r^{2}}=\frac{\alpha_{m}\tilde{p}}{f'(\phi)}-V(\phi)=\frac{\alpha_{m}p}{[f'(\phi)]^{2}}-\frac{\phi
    f'(\phi)-f(\phi)}{2(f'(\phi))^{2}},
\end{equation}
and making algebraic manipulations to separate $\gamma_{,r}$ we
obtain
\begin{equation}\label{eq:28}
\gamma_{,r}=\frac{2\tilde{M}(r)+r^{3}[\alpha_{m}\tilde{p}[f'(\phi)]^{-1}-V(\phi)]}{r(r-2\tilde{M}(r))},
\end{equation}
where $\tilde{M}(r)$ is given by (\ref{eq:20}) and $V(\phi)$ by
(\ref{eq:13}).

The use of a conformal transformation , Einstein gauge, is only used
as a trick which allows us to simplify calculations, namely to
calculate $\gamma_{,r}$.

The matter action must be invariant under diffeomorphisms and the
matter fields satisfy the matter field equations, thus
\cite{Barr96.1}:
\begin{equation}\label{eq:29}
    \nabla^b T_{ab}=0.
\end{equation}
On the other hand, in the Einstein gauge the energy-momentum tensor
is not covariantly conserved \cite{po1}. Therefore, in the Jordan
gauge, a test particle will follow the geodesics of the metric
connection but this is not true in the Einstein gauge.

We assume the Jordan gauge as the physical, in order to research
observational consequence of this hypothesis.

The equation (\ref{eq:29}) is written in Jordan gauge, and the
energy momentum tensor is given by (\ref{eq:25}) since we model the
star as a perfect fluid. Choosing the four-velocity pointing into
the future like direction and normalizing it to $u_{a}u^{a}=-1$, it
becomes
\begin{equation}\label{eq:30}
    u_{\mu}=(\sqrt{A(r')},0,0,0),
\end{equation}
For our metric, in Jordan gauge (\ref{eq:21}), the only nontrivial
component gives,
\begin{equation}\label{eq:31}
    \frac{dp}{dr'}=-\frac{(p+\rho)}{2}g^{ t t}\frac{dg_{ t t}}{dr'},
\end{equation}
  and in order to combine this equation with equation (\ref{eq:28}), we now have
  to consider  equation (\ref{eq:6}), then:
  \begin{equation}\label{eq:32}
    \frac{dp}{dr'}=-\frac{(p+\rho)}{2}\frac{d\gamma}{dr'}+\frac{(p+\rho)}{\sqrt{f'(\phi)}}\frac{d\sqrt{f'(\phi)}}{dr'}.
  \end{equation}

 Making a change by taking into account $r'=\frac{r}{\sqrt{f'(\phi)}}$
 \begin{equation}\label{eq:33}
\frac{d\gamma}{dr'}=(\sqrt{f'(\phi)}+r'\frac{d\sqrt{f'(\phi)}}{dr'})\frac{d\gamma}{dr},
 \end{equation}
and putting everything in terms of $r'$ on the right hand side of
equation (\ref{eq:28}), we can write:
\begin{equation}\label{eq:34}
    \frac{d\gamma}{dr'}=(1+\frac{r'}{\sqrt{f'(\phi)}}\frac{d\sqrt{f'(\phi)}}{dr'}) (\frac{r'^{3}(\alpha_{m}
    p
    [f'(\phi)]^{-1}-f'(\phi)V(\phi))+2\tilde{M}(r')}{r'(r'-2\tilde{M}(r'))}).
\end{equation}

Finally, we are ready to write the equilibrium hydrostatic equation
in the Jordan gauge:
\begin{eqnarray}
-\frac{1}{(p+\rho)}\frac{dp}{dr'}=
\frac{1}{2}(1+\frac{r'}{\sqrt{f'(\phi)}}\frac{d\sqrt{f'(\phi)}}{dr'})\times
\;
\;\;\;\;\;\;\;\;\;\;\;\;\;\;\;\;\;\;\;\;\;\;\;\;\;\,\;\;\;\;\;\;\;\;\;\;\;\;\;\;\;\;\;\;\;\;\;\;\;\;\,\;\;\;\;\;\;\:\;\;\;\;
\nonumber
\\
(\frac{r'^{3}(\alpha_{m} p
[f'(\phi)]^{-1}-f'(\phi)V(\phi))+2\tilde{M}(r')}{r'(r'-2\tilde{M}(r'))})-\frac{1}{\sqrt{f'(\phi)}}\frac{d\sqrt{f'(\phi)}}{dr'}.
\label{eq:35}   \end{eqnarray}

The equation (\ref{eq:35}) can be seen as equivalent to
Tolman-Oppenheimer-Volkoff equation, (for a different approach see
\cite{Kai06.1}). It is easy to show that for General Relativity,
i.e. $f(\phi)=\phi$, the equation (\ref{eq:35}) becomes the known
result. This equation is very difficult to solve, for any model
different from General Relativity, even if we consider an
incompressible fluid.

\section*{V. Newtonian limit of the $f(R)$ gravity}

 In order to obtain the Newtonian limit of the $f(R)$
gravity from the hydrostatic equilibrium equation, we assume
$p\ll\rho$, $Gm(r)\ll r$, where $m(r)=4\pi\int^r_0 \rho(x)x^2dx$,
and take $\rho$ small enough so that it produces a good
approximation in the following Taylor expansion
\begin{equation}\label{eq:36}
    f(\phi(T))\equiv f(T)\approx f(0)- \rho
    \frac{df}{dT}\mid_{T=0}+O(\rho^{2}),
\end{equation}
taking into account $T=3p-\rho \approx -\rho$. Similar expansions
were made for the other quantities appearing in (\ref{eq:35}). In
particular an expansion of $V(\phi)$ is given by the equation
(\ref{eq:13})
\begin{equation}\label{eq:37}
    V(\phi)\approx\frac{f(0)}{2(f'(0))^{2}}-
    \rho(\frac{\alpha_{m}+f'(0)\frac{d\phi}{dT}|_{T=0}}{2(f'(0))^{2}}-\frac{f(0)f''(0)\frac{d\phi}{dT}|_{T=0}}{(f'(0))^{3}})+O(\rho^{2}),
\end{equation}
Thus, taking into account equation (\ref{eq:4}) in order to solve
$\frac{d\phi}{dT}|_{T=0}$, and considering this equation in the case
$T=0$ we can write:
\begin{equation}\label{eq:38}
    V(\phi)\approx\frac{f(0)}{2(f'(0))^{2}}+O(\rho^{2}).
\end{equation}

Now, let us approximate the integral equation (\ref{eq:24}). First
we consider its integrand:

\begin{equation}\label{eq:39}
\frac{\alpha_{m}}{f'(\phi)}T^{t}_{t}-\frac{[\phi
f'(\phi)-f(\phi)]}{2[f'(\phi)]^{2}}\approx
-\frac{\alpha_{m}\rho}{[f'(0)]^{2}}-\frac{f(0)}{2[f'((0)]^{2}}+O(\rho^{2}).
\end{equation}

The integral part of $\tilde{M}(r')$ can be written in the form
 \begin{equation}\label{eq:40}
\int^{r'\sqrt{f'(\phi)}}_{0}\hat{r}^{2}(\frac{\alpha_{m}}{f'(\phi)}T^{t}_{t}-\frac{[\phi
    f'(\phi)-f(\phi)]}{2[f'(\phi)]^{2}})d\hat{r} \approx
    \int^{r'\sqrt{f'(0)}}_{0}(\frac{-\alpha_{m}\rho}{[f'(0)]^{2}})\hat{r}^{2}d\hat{r}+\frac{1}{6}
    \frac{f(0)}{[f'(0)]^{2}}(r'\sqrt{f'(\phi)})^{3},
 \end{equation}
where we have assumed $\rho$ continuum. Finally, the expression
(\ref{eq:24}) takes the form
\begin{equation}\label{eq:41}
    \tilde{M}(r')\approx \frac{G
    m(r')}{f'(0)}(1+\frac{\rho}{f'(0)}\frac{df'}{dT}|_{T=0})+\frac{1}{12}\frac{f(0)}{f'(0)}r'^{3},
\end{equation}
where $m(r')$ is the usual integral expression for the mass. Another
term in the expression (\ref{eq:35}) is
\begin{equation}\label{eq:42}
\alpha_{m}p [f'(\phi)]^{-1}-f'(\phi)V(\phi)\approx
[{\frac{\alpha_{m}p}{f'(0)}(1+\frac{\rho}{f'(0)}\frac{df'}{dT}|_{T=0})+\rho\frac{f(0)}{2[f'(0)]^{2}}\frac{df'}{dT}|_{T=0}}]-\frac{1}{2}\frac{f(0)}{f'((0)},
\end{equation}
The terms in the brackets in (\ref{eq:42}) can be ignored because
has terms of superior order. Plugging the approximations into
equation (\ref{eq:35}) and working consistently with the conditions
in this limit gives

\begin{equation}\label{eq:43}
    -\frac{dp}{dr'}\approx \frac{\rho}{f'(0)}(\frac{G
    m(r')}{r'^{2}})-\frac{\rho}{6}\frac{f(0)}{f'(0)}r'-\frac{\rho}{\sqrt{f'(\phi)}}\frac{d\sqrt{f'(\phi)}}{dr'},
\end{equation}
solving the derivative in the right hand side of equation
(\ref{eq:43}) and making one more approximation keeping to linear
order in $\rho$ we have:
\begin{equation}\label{eq:44}
    -\frac{dp}{dr'}\approx \frac{\rho}{f'(0)}\frac{G
    m(r')}{r'^{2}}-\frac{\rho}{6}\frac{f(0)}{f'(0)}r'+\frac{\rho}{2}\frac{f''(0)}{f'(0)}\frac{\alpha_{m}}{(\phi(0)f''(0)-f'(0))}\frac{d\rho}{dr'}.
\end{equation}

If the particular choice of theory of gravity is such that $f(0)=0$,
we have a theory without cosmological constant, and the last
equation can be written in the form:

\begin{equation}\label{eq:45}
    -\frac{dp}{dr'}\approx \frac{\rho}{f'(0)}\frac{G
    m(r')}{r'^{2}}-\frac{\rho\alpha_{m}}{2}\frac{f''(0)}{[f'(0)]^{2}}(\frac{d\rho}{dr'}).
\end{equation}

The last expressions give the weak field limit from the hydrostatic
equilibrium equation
 for the theory, and they are totaly in accordance with
previous works \cite{Do04.1}. However, the Taylor expansion
(\ref{eq:36}) can be done in general, when $\rho$ is small enough,
but not for singular theories of type $f(R)=R-\frac{a^{2}}{3R}$, in
which the condition in order to keep the linear term in $\rho$ is
$\rho \alpha_{m} \ll a$ \cite{So06.1}, where $a$ is essentially the
vacuum energy density with $\rho_{vac}=a/\alpha_{m}\sim
10^{-31}[gr/cm^{3}]$. It is obvious that the mean density of our
solar system $\rho \sim 10^{-11} [gr/cm^3]$ does not satisfy this
condition, and then we can not use the above approximation.
Therefore, in the next section we will study the weak field limit
for this kind of theories in the case of $\rho$ being of order of
the mean density of our solar system.

\section*{VI. Newtonian limit of the singular $f(R)$ gravity}

Theories like those presented in \cite{Vo03.1}, i.e. $R -
\frac{{a}^2}{R}$, which have a pole in $R=0$, seem to explain the
present accelerated expansion. But,there are still  some obscure
points about how to obtain their behavior in a weak field limit,
particularly what metric we have to use as a background metric.
Dominguez and Barraco \cite{Do04.1} perturbed around the de Sitter
space-time, which is the maximal symmetric vacuum solution of the
theory. Sotiriou \cite{So06.1} claimed that this procedure was
erroneous, when $\rho$ is near the mean density of the solar system,
which, according to the final part of the last section, is correct.
Therefore, he perturbed around the flat metric, which is not a
solution of the theories but it could be near the vacuum solutions.

In our present work we suppose nothing about the background metric
in order to obtain the Newtonian limit, and this is one of the
principal difference with the previous works. First we find this
limit for the theory:
\begin{equation}\label{eq:46}
    f(R)=R-\frac{a^{2}}{3R}.
\end{equation}

For this theory, equation (\ref{eq:4}) gives

\begin{equation}\label{eq:47}
    \phi(T)=\frac{1}{2}(-\alpha_{m}T\pm\sqrt{\alpha_{m}^{2}T^{2}+4a^{2}}),
\end{equation}
 where we have to choose the minus sign in front the square root, in order
 to return to General Relativity when $a\longrightarrow0$. We
 consider, $p \ll \rho$,  $r'\gg Gm(r')$ and:

\begin{equation}\label{eq:48}
    \frac{a}{\alpha_{m}\rho}\ll1.
\end{equation}

Here we are considering \cite{So06.1} $\rho=10^{-11}[gr/cm^{3}]$,
$a\sim10^{-67}[eV]^{2}$ so
$\mid\frac{a}{\alpha_{m}T}\mid\sim10^{-21}$ where $T\approx-\rho$.
Then we can state:

$$\phi(T)\approx \alpha_{m}\rho+(\frac{a}{\alpha_{m}\rho})a,$$

$$f(\phi)\approx \alpha_{m}\rho+\frac{2}{3}(\frac{a}{\alpha_{m}\rho})a,$$

$$f'(\phi)\approx 1+ \frac{1}{3}(\frac{a}{\alpha_{m}\rho})^{2},$$

\begin{equation}\label{eq:49}
f''(\phi)\approx
-\frac{2}{3}(\frac{a}{\alpha_{m}\rho})^{2}\frac{1}{\alpha_{m}\rho}.
\end{equation}

 Next, we have to approximate the others quantities appearing in the
 hydrostatic equilibrium equation, and take into account that in this
approximation $\tilde{M}(r')\approx Gm(r')$, where $m(r')$ is the
ordinary mass appearing  in General Relativity. Then, plugging into
(\ref{eq:35}) we have,
$$-\frac{dp}{dr'}=\rho\frac{Gm(r')}{r'^{2}}-\frac{1}{3}(\frac{a}
{\alpha_{m}\rho})^{2}\frac{dT}{dr'}(1-\frac{Gm(r')}{r'})$$
\begin{equation}\label{eq:50}
     \approx
     \rho\frac{Gm(r')}{r'^{2}}-\frac{1}{3}(\frac{a}{\alpha_{m}\rho})^{2}\frac{d(3p-\rho)}{dr'},
\end{equation}
 we have already neglected the pressure but we can't neglect its
derivative. Therefore, the second line is obtained from the first by
substituting $T=3p-\rho$. Using (\ref{eq:48}) we have, in the weak
field approximation:
\begin{equation}\label{eq:51}
    -\frac{dp}{dr'}\approx\rho\frac{Gm(r')}{r'^{2}}+\frac{1}{3}(\frac{a}{\alpha_{m}\rho})^{2}\frac{d\rho}{dr'},
\end{equation}
 where the terms of order $a^2$ are kept only in gradients of
$\rho$, where they are relevant. The extra term in equation
(\ref{eq:51}) is the correction of lowest order in
$\frac{a}{\alpha_{m}\rho}$  to the purely Newtonian behavior.
Therefore, the last equation gives a good Newtonian limit for the
theory. This limit is in accordance with the result obtained by
Sotiriou.

We can show more generally that the theories
\begin{equation}\label{eq:52}
    f(R)=R-\frac{1}{n+2}\frac{a^{n+1}}{R^{n}},
\end{equation}
also has a good Newtonian limit. Equation (\ref{eq:4}) for these
theories give
\begin{equation}\label{eq:53}
\phi^{n+1}+\alpha_{m}T\phi^{n}-a^{n+1}=0,
\end{equation}
where the last equation has an approximate solution of the form
$\phi(T)\approx-\alpha_{m}T\approx\alpha_{m}\rho$. Then using
the expressions
$$ f(\phi)\approx\alpha_{m}\rho+\frac{2}{n+2}(\frac{a}{\alpha_{m}\rho}^n)a,$$
$$f'(\phi)\approx 1+\frac{n}{n+2}(\frac{a}{\alpha_{m}\rho})^{n+1},$$
\begin{equation}\label{eq:54}
    f''(\phi)\approx-
\frac{n(n+1)}{n+2}(\frac{a}{\alpha_{m}\rho})^{n+1}\frac{1}{\alpha_{m}\rho},
\end{equation}
 considering these approximations in the context of equation (\ref{eq:35})
and carrying out exactly the same procedure as before:
\begin{equation}\label{eq:55}
    -\frac{dp}{dr'}\approx \rho\frac{Gm(r')}{r'^{2}}+
    \frac{n}{2}\frac{(n+1)}{(n+2)}(\frac{a}{\alpha_{m}\rho})^{n+1}\frac{d\rho}{dr'}.
\end{equation}

Where the last term is the correction of lowest order in
$\frac{a}{\alpha_{m}\rho}$. Note that for the case $n=1$, the last
equation returns to equation (\ref{eq:51}) and for $n=0$ we return
to General Relativity

Finally, the above equation shows that these theories have also a good Newtonian limit.

\section*{VII. Conclusions}

 We have obtained the hydrostatic equilibrium equation in the Jordan
 gauge, where the matter obeys simple conservation laws.
 This equation is the generalization of  the Tolman-Oppenheimer-Volkoff equation in General Relativity.

We have studied the Newtonian Limit, in this gauge, from the hydrostatic equilibrium equation, and in section V we found an expression in the weak field limit for the pressure gradient. In this section we saw that this
result is not correct if we consider $f(R)=R-\frac{a^{2}}{3R} $
gravity models with density typically related to a Newtonian regime.
However the expansion (\ref{eq:45}) for $\frac{dp}{dr'}$ is
perfectly accurate for this singular gravity if the conditions are
such that $\alpha_{m}\rho \ll a$. And, if this is the case, the
theory does not give a good Newtonian Limit as previously shown
Barraco and Dominguez \cite{Do04.1}.

In section VI we have shown that the singular theory,
$f(R)=R-\frac{a^{2}}{3R}$, and its generalization,
$f(R)=R-\frac{1}{n+2}\frac{a^{n+1}}{R^{n}}$, have  good Newtonian
limits for densities such as the mean density of the solar system.
The difference with previous works \cite{Do04.1,So06.1} is that we
found the Newtonian limit, in the spherically symmetric case, with
no assumption about the background metric. Finally, in the general
case, we have shown that $n$ controls the departure from purely
Newtonian behavior.
\section*{ACKNOWLEDGMENTS}
The authors are very grateful to CONICET and SeCYT, Argentina, for
financial support.


\begin{thebibliography}{99}
\bibitem{Ri98.1} A. G. Riess \emph{et al.}, {\sl Astron. J.}{\bf116},1009 (1998).
\bibitem{Pe99.1} S. Perlmutter \emph{et al.}, {\sl Astrophys.
J.} {\bf517},565 (1999).
\bibitem{Be03.1} C. L. Bennet \emph{et al.},{\sl Astrophys. J.
(Suppl.)} {\bf148},1 (2003).
\bibitem{Ne02.1} C. B. Netterfeld \emph{et al.}, {\sl Astrophys.
J.}{\bf571},604 (2002).
\bibitem{Ca02.1} S. M. Capozzielo \emph{et al.}, {\sl Int. J. Mod.
Phys.} {\bf D 11}, 604 (2002).
\bibitem{No03.1} S. Nojiri and S. D. Odinstov \emph{et al.}, {\sl Phys.
Rev.} {\bf D 68}, 123512 (2003).
\bibitem{Do03.1} A. D. Dogolov and M. Kawasaki \emph{et al.}, {\sl Phys. Lett.} {\bf B 573}, 1 (2003).
\bibitem{Do6.10.06} S.N. Nojiri and S.D. Odintsov {\sl
arXiv:hep-th0310045} (2006).
\bibitem{Od06.1} S. Nojiri and S. D. Odinstov {\sl
arXiv:gr-qc/0601213 v5} (2006).
\bibitem{Di04.1} R. Dick, {\sl Gen. Relativ.
Gravit.}, {\bf36}, 217 (2004).
\bibitem{Ha93.1} V. H. Hamity and D. E. Barraco, {\sl Gen. Relativ.
Gravit.}, {\bf25}, 461 (1993).
\bibitem{Barr91.1} Barraco D.E.,
Ph.D. Thesis, Universidad Nacional de C\'{o}rdoba (1991).
\bibitem{Fran93.1} Ferraris M., Francaviglia M. and Volovich I., Il
Nuovo Cimento {\bf 108B, N. 11} 1313 (Nov. 1993).
\bibitem{Fran94.1} Ferraris M., Francaviglia M. and Volovich I.,
Class. Quantum Grav. {\bf 11}, 1505 (1994).
\bibitem{Barr99.1} Barraco D., Dominguez E., Guibert R., Phys. Rev. D {\bf
60}, 44012 (1999).
\bibitem{Barr00.1} Barraco D., Hamity V.H., Phys. Rev. D {\bf 62},
044027 (2000).
\bibitem{Barr96.1} D. E. Barraco, R. Guibert , V. H. Hamity, and H.
Vucetich, {\sl Gen. Relativ. Gravit.}, {\bf28}, 339 (1996).
\bibitem{Barr02.1} D. E. Barraco, V. H. Hamity and H. Vucetich, {\sl Gen. Relativ.
Grav.} {\bf34}, 533 (2002).
\bibitem{Vo03.1} D. N. Vollick, {\sl Phys. Rev.}, {\bf D 68},
063510 (2003).
\bibitem{Od06} G. Allemandi, A. Borowiec, M.
Francaviglia, and S. Odintsov {\sl arXiv:gr-qc/0504057 v2} (2006).
\bibitem{Ma94.1} G. Magnano and L. M. Sokolowski, {\sl Phys.
Rev.} {\bf D 50}, 5039 (1994).
\bibitem{So06.1} Thomas P. Sotiriou {\sl arXiv:gr-qc/0507027
v2}(2006).
\bibitem{Do04.1} A. E. Dominguez and D. E. Barraco, {\sl Phys.
Rev} {\bf70}, 043505 (2004).
\bibitem{bust1} A. Bustelo, {\sl
Graduation Thesis, Universidad Nacional de C\'ordoba}, September
2006.
\bibitem{Po06.1} N. J. Poplawski, {\sl Class. Quantum Grav.} {\bf23}, 2011 (2006).
\bibitem{Od05.1} G. Cognola, E. Elizalde, S. Nojiri, S. Odintsov and
S. Zerbini {\sl arXiv: hep-th/0501096} (2005).
\bibitem{Sy64.1} J. L. Synge, {\sl Relativity: The General Relativity}(North- Holland, Amsterdam,
1976).
\bibitem{po1} N. J. Poplawski {\sl Class. Quantum Grav.} {\bf23},
4819 (2006).
\bibitem{Kai06.1} K. Kainulainen, V. Reijonen, and D. Sunhede {\sl
arXiv:qr-qc/0611132} (2006).




\end{thebibliography}
\end{document}